\documentclass[prd,reprint,nofootinbib,superscriptaddress,preprintnumbers,showpacs]{revtex4-1}
\usepackage{amsfonts}
\usepackage{amssymb}
\usepackage{amsmath}
\usepackage{graphicx,color}
\usepackage{dcolumn}
\usepackage{epsfig}
\usepackage{bm}
\usepackage{bbm}
\usepackage{ulem}
\usepackage{hyperref}
\usepackage{lineno}
\usepackage{float}
\usepackage{subfigure}

\allowdisplaybreaks[4]

\begin{document}
	
	\title{Pole determination of $X(3960)$ and $X_0(4140)$ in decay $B^+\to K^+D_s^+D_s^-$}
	\author{Jialiang Lu}
	\affiliation{School of Physics and Optoelectronics Engineering, Anhui University, Hefei 230601, People's Republic of China }
	\author{Xuan Luo}	
	\email{xuanluo@ahu.edu.cn}
    \affiliation{School of Physics and Optoelectronics Engineering, Anhui University, Hefei 230601, People's Republic of China }
	\author{Mao Song}
	\affiliation{School of Physics and Optoelectronics Engineering, Anhui University, Hefei 230601, People's Republic of China }
	\author{Gang Li}
	\affiliation{School of Physics and Optoelectronics Engineering, Anhui University, Hefei 230601, People's Republic of China }	
	\begin{abstract} 
		\vspace{0.5cm}

Two near-threshold peaking structures with spin-parities $J^{PC}=0^{++}$ were recently discovered by the LHCb Collaboration in the $D_s^+D_s^-$ invariant mass distribution of the decay process $B^+\to D_s^+D_s^-K^+$. In our study, we employed a coupled-channel model to fit the experimental results published by the LHCb collaboration, simultaneously fitting the model to the invariant mass distributions of $M_{D_s^+D_s^-}$, $M_{D_s^+K^+}$, and $M_{D_s^-K^+}$. We utilized a coupled-channel model to search for the poles of $X(3960)$ and $X_0(4140)$.  The determination of the poles is meaningful in itself, and it also lays an foundation for the future research on $X(3960)$ and $X_0(4140)$. Upon turning off the coupled-channel and performing another fit, we observed a change in the fitting quality, the effect was almost entirely due to the peak of $X(3960)$, so we suggest that $X(3960)$ may not be a kinematic effect.

	\end{abstract}
	\maketitle	
	\section{Introduction}
	\label{I}
For a long time, we have been asking the question what kind of matter can be formed by quark models. The traditional quark model successfully explains that baryons are complexes of three quarks and mesons are combinations of quarks and antiquarks. With the development of experiments, the recent discovery of candidates for the pentaquark and tetraquark states in experiments has  expanded the scope of our study of traditional hadrons, which include qualitatively different $qqq\bar{q}\bar{q}$, $qq\bar{q}\bar{q}$, in addition, more exotic structures have been observed in experiments; see ref.~\cite{Chen:2016qju,Hosaka:2016pey,Lebed:2016hpi,Esposito:2016noz,Ali:2017jda,Guo:2017jvc,Olsen:2017bmm,Brambilla:2019esw}. To answer the appeal question, we need to determine whether the pentaquark and tetraquark states exist.\\

In order to find the strange state of QCD, the decay process of $B$ mesons will be an important and effective platform. In this process, many candidates for strange hadron states can be observed. In the past few years, major laboratories have successively discovered candidates for strange hadron states in the decay of $B$ mesons, such as $Z_{cs}(4000)$ and $Z_{cs}(4000)$~\cite{LHCb:2021uow}, $X(4140)$~\cite{CDF:2009jgo,D0:2013jvp} in $B^+\to J/\psi\phi K^+$, as well as $X_0(2900)$ and $X_1(2900)$ in $B^+\to D^+D^-K^+$ decay~\cite{LHCb:2020bls,LHCb:2020pxc}. Referring to these experiments, we can find that the three-body decay of $B$ mesons can provide a lot of information about hadron resonance; see ref.~\cite{Xing:2022uqu,Duan:2023qsg,Lyu:2023jos,Li:2023nsw}.\\ 

Very recently, the LHCb Collaboration reported a new near-threshold structure named $X(3960)$ in the $D_s^+D_s^-$ invariant mass distribution of the decay $B^+\to D_s^+D_s^-K^+$. The peak structure is very close to the $D_s^+D_s^-$ threshold with a statistical significance larger than 12$\sigma$. The mass, width, and quantum numbers of this structure were measured to be: $M=3956\pm5\pm10$ MeV and $\Gamma=43\pm13\pm8$ MeV, $J^{PC}=0^{++}$. The LHCb analysis indicates that this structure is an exotic candidate consisting of $cs\bar{c}\bar{s}$ constituents. In addition, when checking the data of the $D_s^+D_s^-$ invariant mass distribution, a dip around $4.14$ GeV can be found, the LHCb interpreted it as another structure named $X_0(4140)$ with mass $M=4133\pm6\pm6$ MeV, width $\Gamma=67\pm17\pm7$ MeV, and quantum numbers $J^{PC}=0^{++}$~\cite{LHCb:2022aki}. As analyzed by the LHCb collaboration, the $X_0(4140)$ might be caused either by a new resonance with the $0^{++}$ assignment or by a $D_s^+D_s^--J/\psi\phi$ coupled-channel effect, but no firm conclusion has been reached there~\cite{LHCb:2022aki}.\\

There are many theoretical studies that have shown great interest in X resonances. In recent years, many articles have used different models and technical methods to study the characteristics of exotic mesons $cs\bar{c}\bar{s}$~\cite{Nieves:2012tt,Wang:2013exa,Lebed:2016yvr,Chen:2017dpy,Meng:2020cbk,Agaev:2017foq,Sundu:2018toi,Agaev:2022iha}. To figure out the origin and structure of $X(3960)$ in decay $B^+\to D_s^+D_s^-K^+$, many explanations have been put forward for the possibility of this structure. Since its mass is close to the $D_s^+D_s^-$ threshold, this structure can be interpreted as the possibility of hadronic molecular. In ref.~\cite{Mutuk:2022ckn,Xin:2022bzt}, it was proposed to treat $X(3960)$ as the molecular state of $D_s^+D_s^-$ with $J^{PC}=0^{++}$ in the QCD sum rules approach. Another calculation with QCD two-point sum rules~\cite{Agaev:2022pis} leads to the assignment that the $X(3960)$ is a scalar diquark-antidiquark state. The calculations in the one-boson-exchange model~\cite{Chen:2022dad} also favor the molecule interpretation. It can also be analyzed by the nature of $X(3960)$ by the coupled-channel method. The authors of Ref.~\cite{Bayar:2022dqa} performed a coupled-channel calcuation of the interaction $D\bar{D}-D_s^+D_s^-$ in the chiral unitary approach and interpreted $X(3960)$ as a hadronic molecule in the coupled $D\bar{D}-D_s^+D_s^-$ system~\cite{Bayar:2022dqa,Ji:2022uie,Ji:2022vdj}. The author of Ref.~\cite{Badalian:2023qyi} interpreted $X(3960)$ as $cs\bar{c}\bar{s}$ state, while in Ref.~\cite{Guo:2022crh}, the $X(3960)$ is interpreted as $0^{++}$ $cs\bar{c}\bar{s}$ tetraquark states by an improved chromomagnetic interaction model. In addition, it was suggested that the $X(3960)$ has probably the mixed nature of a $c\bar{c}$ confining state and $D_s\bar{D}_s$ continuum~\cite{Chen:2023eix}. There has also been some theoretical and experimental work on $X_0(4140)$, but its origins are still debated. For instance, in Ref.~\cite{Guo:2022crh}, the $X_0(4140)$ is also interpreted as $cs\bar{c}\bar{s}$ tetraquak states. The discussion about mass and width in Ref.~\cite{Agaev:2022pis} allowed us to believe that the  model is also acceptable. Since different computational models suggested different explanations, coming up with new ideas and insights into this state can go a long way in helping us shed more light on the origin of $X_0(4140)$.\\

In this study, we analyze the decay process $B^+\to D_s^+D_s^-K^+$ as published by the LHCb collaboration. We simulated a coupled-channel model to analyze the data~\cite{Nakamura:2022gtu}, using the default model and fitting the $M_{D_s^+D_s^-}$, $M_{D_s^+K^+}$, and $M_{D_s^-K^+}$ of these three different invariant mass distributions. Using the amplitude provided by the coupled channel model, we address the following questions: (i) the pole position of $X(3960)$ and $X_0(4140)$; and (ii) whether the production of $X(3960)$ is solely due to a kinematic effect.
 
	\section{Framework}
	\label{II}
\begin{figure*}[htpb]
	\centering
	\includegraphics[width=0.32\textwidth]{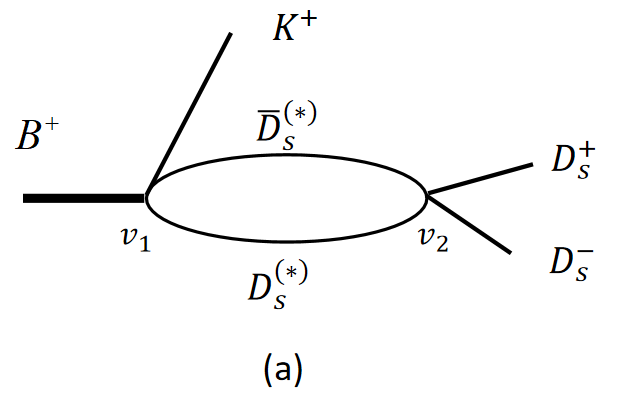}
	\includegraphics[width=0.2\textwidth]{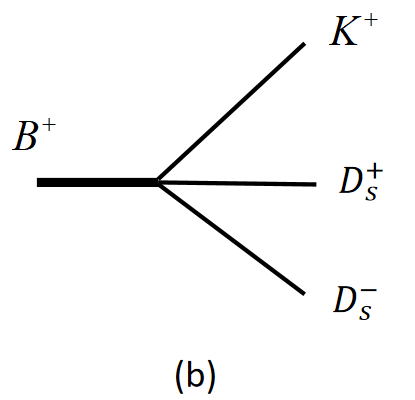}
	\includegraphics[width=0.26\textwidth]{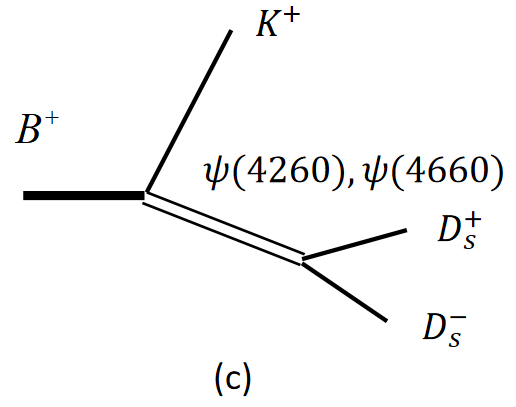}
	\caption{Contributions of three mechanisms in decay $B^+\to K^+D_s^+D_s^-$. (a) Coupled-channel; (b) Direct production; (c) Breit-Weigner effects. }
	\label{fig1}
\end{figure*}

The LHCb data shows visible structures $X(3960)$ and $X_0(4140)$ around the $D_s\bar{D}_s$, $D_s^{\ast}\bar{D}_s^{\ast}$ thresholds respectively. Thus it is reasonable to assume that the structures are caused by the threshold cusps that are further enhanced or suppressed by hadronic rescatterings and the associated poles~\cite{Dong:2020hxe,Nakamura:2022gtu}; see Fig.~\ref{fig1}(a). Meanwhile, for the two peaks at 4260 MeV and 4660 MeV, we refer to the suggestions given in the LHCb and add two Breit-Weigner effects, as shown in Fig.~\ref{fig1}(c). We assume that other possible mechanisms are absorbed by the direct decay mechanism in Fig.~\ref{fig1}(b).\\

First of all, we present amplitude for Fig.~\ref{fig1}(a). The first vertex $v_1$ is a weak interaction, the initial weak $B^+\to K^+D_s\bar D_s$ vertex is
\begin{align}
	v_1=c_{\alpha,B^+K^+}f_{D_s\bar{D}_s}^0F_{K^+B^+}^0,
	\label{v1(1)}
\end{align}
and the initial weak $B^+\to K^+D_s^{\ast}\bar D_s^{\ast}$ vertex is
\begin{align}
	v_1=c_{\alpha,B^+K^+}\vec{p}_{K^+}\cdot(\vec{\epsilon}_{D_s^*}\times\vec{\epsilon}_{\bar{D}_s^*})f_{D_s^*\bar{D}_s^*}^0F_{K^+B^+}^0.
	\label{v1(2)}
\end{align}
The energy, momentum, and polarization vector of a particle $x$ are denoted by $E_x$, $\bm{p}_x$, and $\bm{\epsilon}_x$, respectively, and particle masses are from Ref.~\cite{ParticleDataGroup:2020ssz}. The  $c_{\alpha,B^+K+}$ is a complex coupling constant, which reprensnt $c_{D_s\bar{D}_s,B^+K+}$ and $c_{D_s^{\ast}\bar{D}_s^{\ast},B^+K+}$. We introduced form factors $f_{ij}^L$ and $F_{kl}^L$ defined by
\begin{align}
	f_{ij}^L=\dfrac{1}{\sqrt{E_iE_j}}\left(\dfrac{\Lambda^2}{\Lambda^2+q_{ij}^2}\right)^{2+\frac{L}{2}},\\ F_{kl}^L=\dfrac{1}{\sqrt{E_kE_l}}\left(\dfrac{\Lambda^2}{\Lambda^2+\tilde{p}_k^2}\right)^{2+\frac{L}{2}}.
	\label{form factor}
\end{align}
\\
where $q_{ij}$ is the momentum of $i$ in the $ij$ center-of-mass frame, and the $\tilde{p}_{k}$ is the momentum of $k$ in the total center-of-mass frame.The $\Lambda$ is a cutoff, and $\Lambda=1$~GeV; for all the interaction vertices, this cutoff value is the same.\\

The second vertex $v_2$ is hadron scattering, the perturbative interactions for $D_s\bar{D}_s(D_s^{\ast}\bar{D}_s^{\ast})\to D_s\bar{D}_s$ are given by $s$-wave separable interactions. For $D_s\bar{D}_s\to D_s\bar{D}_s$,
\begin{align}
	v_2=h_{D_s^+D_s^-,D_s\bar{D}_s}f_{D_s^+D_s^-}^0f_{D_s\bar{D}_s}^0, 
	\label{v2(1)}
\end{align}
and for $D_s^{\ast}\bar{D}_s^{\ast}\to D_s\bar{D}_s$,
\begin{align}
	v_2=h_{D_s^+D_s^-,D_s^*\bar{D}_s^*}\vec{\epsilon}_{D_s^*}\cdot\vec{\epsilon}_{\bar{D}_s^*}f_{D_s^+D_s^-}^0f_{D_s^*\bar{D}_s^*}^0.
\end{align}

There is also a vertex between the two vertices, which is the coupling of the two loops, which we call $v_3$. The coupling of different loops is similar in form, we consider the coupled-channel: $D_s^{\ast}\bar{D}_s^{\ast}-D_s\bar{D}_s$\\
\begin{align}
	v_3=\vec{\epsilon}_{D_s^*}\cdot\vec{\epsilon}_{\bar{D}_s^*}G_{D_s^*\bar{D}_s^*,D_s\bar{D}_s}(M_{D_s^+D_s^-})
\end{align}
We introduce $G_{\beta\alpha}(E)=[\delta_{\beta\alpha}-h_{\beta,\alpha}\sigma_\alpha(E)]^{-1}$, $h_{\beta,\alpha}$ is a coupling constant, where $\alpha$ and $\beta$ label interaction channels,  with
\begin{align}
	&\sigma_{D_s\bar{D}_s}(E)=\int dqq^2\dfrac{[f_{D_s\bar{D}_s}^0(q)]^2}{E-E_{D_s}(q)-E_{\bar{D}_s}(q)+i\varepsilon},\\
	&\sigma_{D_s^*\bar{D}_s^*}(E)=\int dqq^2\dfrac{[f_{D_s^*\bar{D}_s^*}^0(q)]^2}{E-E_{D_s^*}(q)-E_{\bar{D}_s^*}(q)+i\varepsilon}.
\end{align}
With the above ingredients, the amplitudes for the fig.~\ref{fig1}(a) are respectively given\\
\begin{align}
	A=
	&4\pi f_{D_s^+D_s^-}^0(p_{D_s^+})F_{K^+B^+}^0\sum_{\alpha}^{D_s\bar{D}_s,D_s^*\bar{D}_s^*}\sum_{\beta}^{D_s\bar{D}_s,D_s^*\bar{D}_s^*}\notag\\
	&c_{\alpha,B^+K^+}G_{\beta\alpha}(M_{D_s^+D_s^-})h_{D_s^+D_s^-,\beta}\sigma_\beta.
	\label{ampu1}	
\end{align}

Regarding the direct decay mechanism of Fig.~\ref{fig1}(b),
\begin{align}
	A_{\rm{dir}}=&c_{\rm{dir}}f_{D_s^+D_s^-}^0F_{K^+B^+}^0.
	\label{ampu2}
\end{align}
with a coupling constant $c_{\rm dir}$.\\
\\

Finally, we consider the Breit-Weigner mechanism of Fig.~\ref{fig1}(c),

\begin{align}
	&A_{\psi(4260)}^{1^-}=c_1\dfrac{\vec{p}_{K^+}\cdot\vec{p}_{D_s^+}f_{D_s^+D_s^-,\psi}^1f_{\psi K^+,B^+}^1}{E-E_{K^+}-E_\psi+\frac{i}{2}\Gamma_{\psi(4260)}},\\
	&A_{\psi(4660)}^{1^-}=c_2\dfrac{\vec{p}_{K^+}\cdot\vec{p}_{D_s^+}f_{D_s^+D_s^-,\psi}^1f_{\psi K^+,B^+}^1}{E-E_{K^+}-E_\psi+\frac{i}{2}\Gamma_{\psi(4660)}}.	
\end{align}
where, the $\vec{p}_{K^+}$ in the $B^+$ CM, the $\vec{p}_{D_s^+}$ in the $D_s^+D_s^-$ CM, the form factor defined by

	\begin{align}
		&f_{D_s^+D_s^-,\psi}^1=\frac{1}{\sqrt{E_{D_s^+}E_{D_s^-}m_\psi}}\left(\frac{\Lambda^2}{\Lambda^2+q_{D_s^+D_s^-}^2}\right)^{\frac{5}{2}},\\
		&f_{\psi K^+,B^+}^1=\frac{1}{\sqrt{E_\psi E_{K^+}E}}\left(\frac{\Lambda^2}{\Lambda^2+q_{\psi K^+}^2}\right)^{\frac{5}{2}}.
	\end{align}
with constant $c_1$ and $c_2$.

	\section{Results}
	\label{III}
\begin{figure*}
	\centering 
	\subfigure[]{\includegraphics[width=0.45\textwidth]{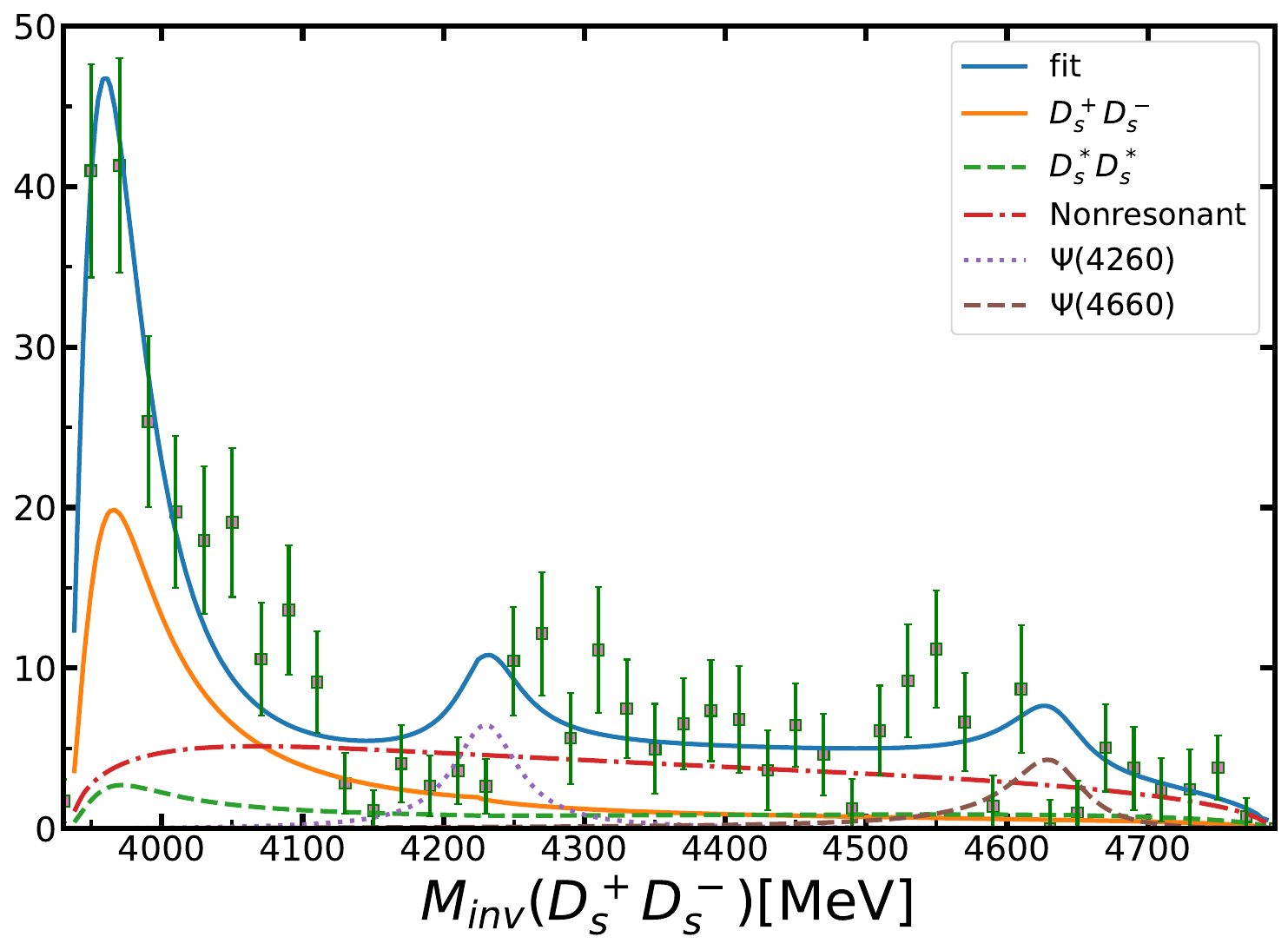}}
	\subfigure[]{\includegraphics[width=0.45\textwidth]{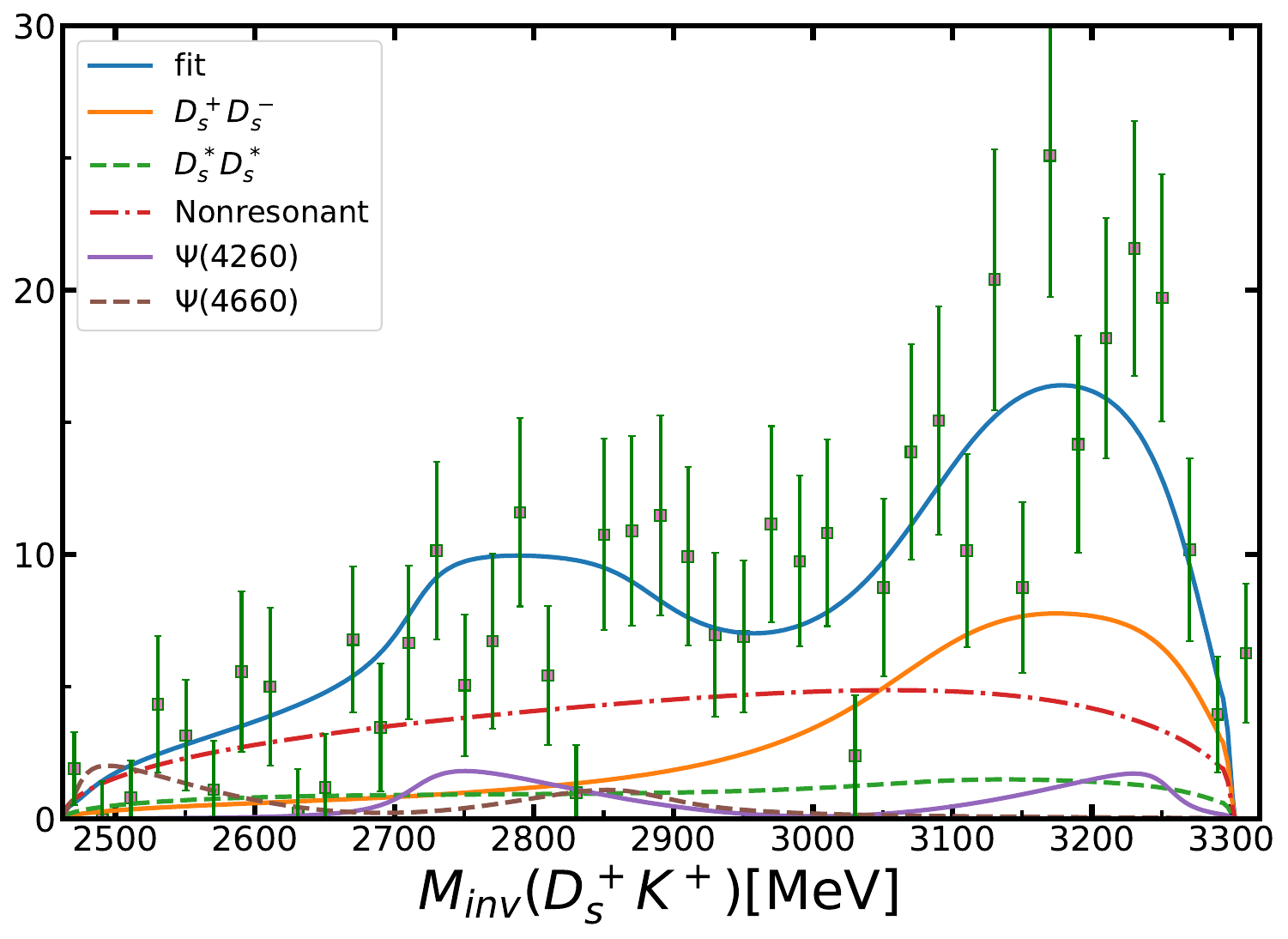}}
	\subfigure[]{\includegraphics[width=0.45\textwidth]{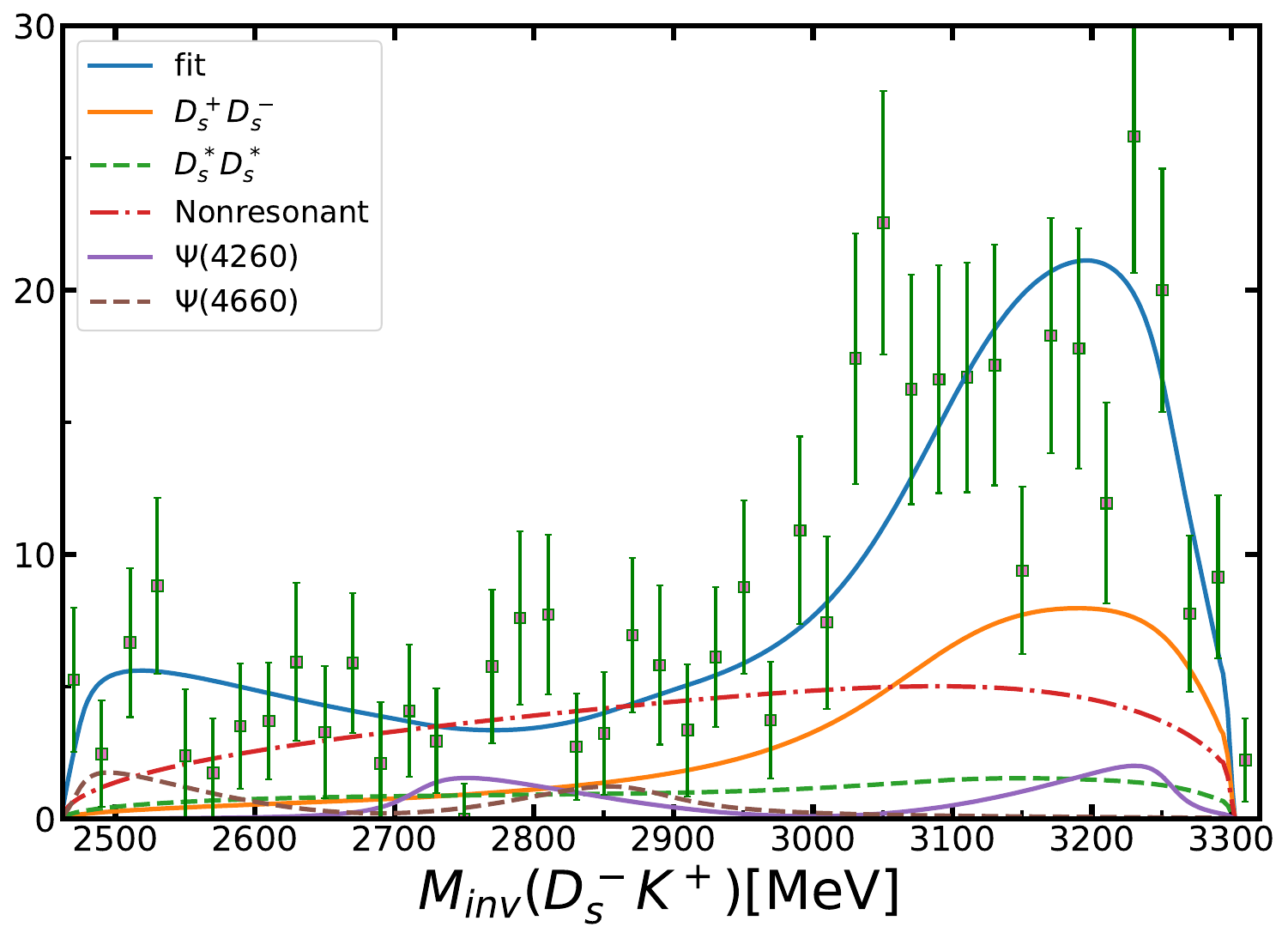}}
	
	\caption{(a)$D_s^+D_s^-$, (b)$D_s^+K^+$, (c)$D_s^-K^+$ invariant mass distribution for $B^+\to D_s^+D_s^-K^+$.} 
	\label{fig2} 
\end{figure*}

\begin{figure*}
	\centering 
	\subfigure[]{\includegraphics[width=0.45\textwidth]{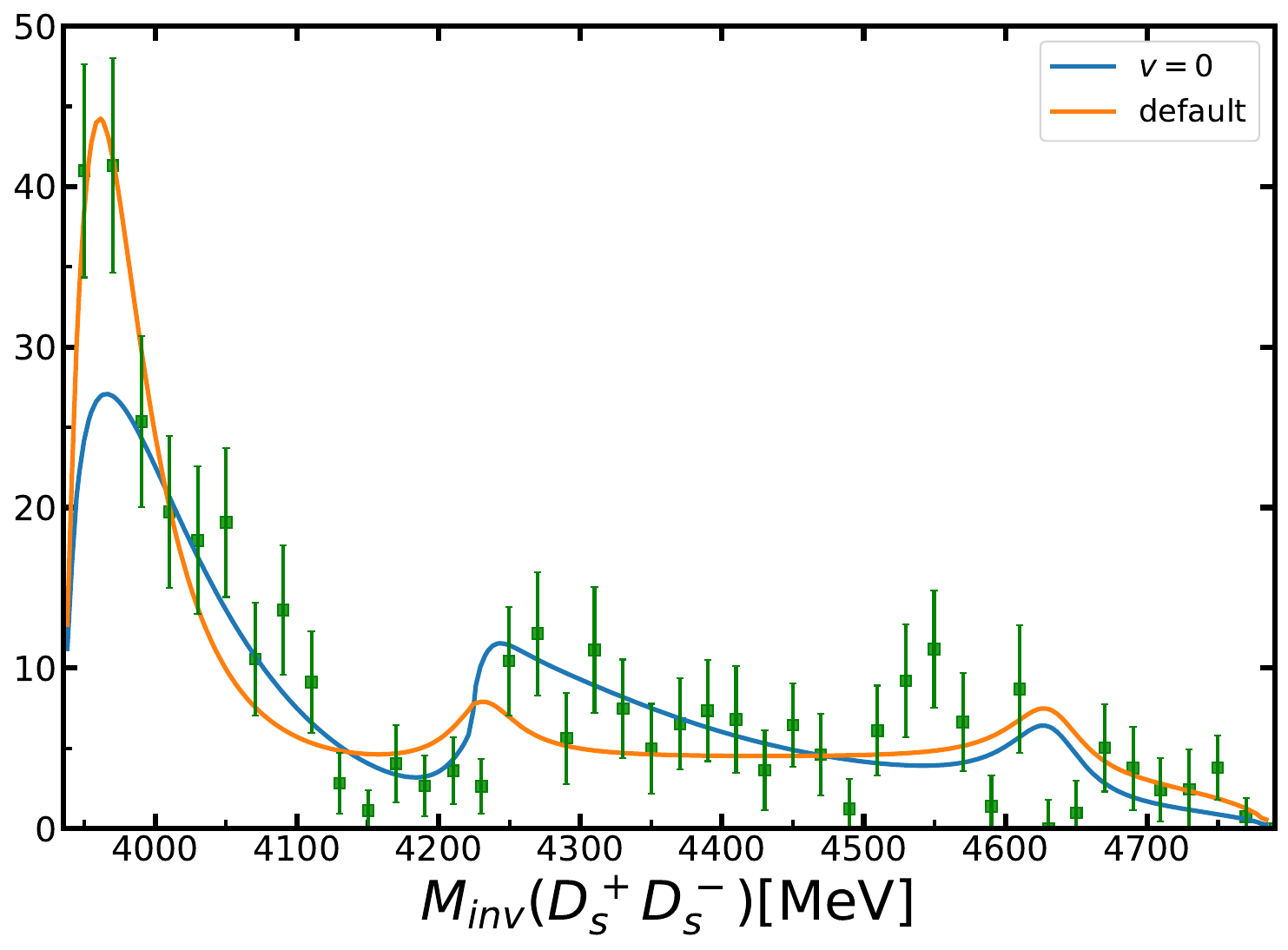}}	
	\caption{Comparison of different models. The orange line in the figure is the default model, while the blue line is the model that turn off the couple-channel effect.} 
	\label{} 
\end{figure*}

\begin{table*}
	\renewcommand{\arraystretch}{1.6}
	\tabcolsep=3.mm
	\caption{Parameter values for $B^+\to D_s^+D_s^-K^+$ models. The second and third columns are for the default and no couple-channel effect models.}
	\begin{tabular}{lcc}
		$c_{D_s\bar{D}_s,B^+K+}$ &0.50+0.28$i$ &-0.49+1.64$i$\\
		$c_{D_s^{\ast}\bar{D}_s^{\ast},B^+K+}$ &0.53-0.28$i$ &-0.03-0.23$i$\\
		$c_{\rm dir}$ &-2.83-60.7$i$ &4.13-7.00$i$\\
		$c_1$ &2.57-3.24$i$ &-3.98-3.04$i$\\
		$c_2$ &-7.41+2.03$i$ &4.81+8.78$i$\\
		$h_{D_s\bar{D}_s,D_s\bar{D}_s}$ &15.61+6.94$i$ &3.13+1.68$i$\\
		$h_{D_s^{\ast}\bar{D}_s^{\ast},D_s^{\ast}\bar{D}_s^{\ast}}$ &23.82-35.67$i$ &0\\
		$h_{D_s^{\ast}\bar{D}_s^{\ast},D_s\bar{D}_s}$&-35.90+9.35$i$ &-15.06+4.62$i$\\
		$\Lambda$ (MeV) &1000 (fixed)&1000 (fixed)\\
	\end{tabular}
\label{table1}
\end{table*}

\begin{table}[b]
	\renewcommand{\arraystretch}{1.6}
	\tabcolsep=2.mm
	\caption{$X(3960)$ and $X_0(4140)$ poles in default model. Pole positions (in MeV) and their Riemann sheets (see the text for notation) are given in the second and third columns, respectively.}
	\begin{tabular}{lcc}
		$X(3960)$ &3941.79+42.12$i$ &$(pu)$\\
		$X_0(4140)$ &4143.44 &$(pp)$\\
	\end{tabular}
\label{table2}
\end{table}

We simultaneously fit the invariant mass distributions of $M_{D_s^+D_s^-}$, $M_{D_s^+K^+}$, and $M_{D_s^-K^+}$ from the LHCb Collaboration using the amplitudes of  Eqs.~\ref{ampu1}. The amplitude includes the vertices of the weak interaction and the adjustable coupling constant brought about by the hadron interaction; this includes $c_{D_s\bar{D}_s,B^+K^+}$,  $c_{D_s^{\ast}\bar{D}_s^{\ast},B^+K^+}$, $c_{\rm dir}$, $c_1$, $c_2$, $h_{D_s\bar{D}_s,D_s\bar{D}_s}$, $h_{D_s\bar{D}_s,D_s^{\ast}\bar{D}_s^{\ast}}$, $h_{D_s^{\ast}\bar{D}_s^{\ast},D_s\bar{D}_s}$, and $h_{D_s^{\ast}\bar{D}_s^{\ast},D_s^{\ast}\bar{D}_s^{\ast}}$. To reduce the number of fitting parameters, we set $h_{D_s\bar{D}_s,D_s^{\ast}\bar{D}_s^{\ast}}=h_{D_s^{\ast}\bar{D}_s^{\ast},D_s\bar{D}_s}$ as allowing them to be different does not significantly affect the quality of the fit. Consider that the coupling constant and the interaction constant of hadron scattering are consistent, we can further reduce the fitting parameters. Finally, given that the magnitude and phase of the full amplitude are arbitrary, our default model has a total of nine fitting parameters. Our default model has a total of 9 (8+1) fitting parameters.\\ 

We show the default model by the solid blue curves in Fig.~\ref{fig2}, which closely matches the LHCb data, we can clearly see the peak at 3960 MeV and a dip at 4140 MeV. The fitting quality is $\chi^2$/ndf=(58+54+56)/(127-9)$\simeq$1.42, where three $\chi^2$ come from three different distributions; "ndf" is the number of bins (43 for $D_S^+D_s^-$ 42 for $D_s^+K^+$ and 42 for $D_s^-K^+$) subtracted by the number of fitting parameters.\\

We also show the different contributions of the chart in Fig.~\ref{fig2}. The solid orange curves represent the contribution of $D_s^+D_s^-$ single channel, and likewise, the dotted green curves represent the contribution of $D_s^{\ast}\bar D_s^{\ast}$ single channel. In general, it is evident that the solid orange curves plays a dominant role throughout the entire process, particularly in relation to the peak of $X(3960)$. This behavior can be attributed to the fact that the $X(3960)$ peak primarily arises from the threshold of $D_s^+D_s^-$. For the peak at 4260 MeV and the peak at 4660 MeV, we adopted the same method as the LHCb Collaboration, and introduced two Breit-Weigner effects~\cite{Cao:2020vab,Cao:2019wwt}, $\psi(4260)$ and $\psi(4660)$, which are represented by purple dotted curves and brown dotted curves, respectively. The analysis here is generally consistent with the analysis given by LHCb; for two peaks near 4260 MeV and 4660 MeV, the final fitting results have been significantly improved.\\

In our study, we conducted a search for poles in the default $D_s\bar D_s-D_s^{\ast}\bar D_s^{\ast}$ coupled-channel scattering amplitude using the method of the analytic continuation. We found the poles of $X(3960)$ and $X_0(4140)$ which are summarized in the table~\ref{table2}. Additionally, in the table, we also list the Riemann sheets of the poles by $(D_s\bar D_s^-D_s^{\ast}\bar D_s^{\ast})$ where $s_\alpha=p$ indicates that the pole is located on the physical $p$ sheet of the channel, while $s_{\alpha}=u$ indicates that it is on the unphysical $u$ sheet of the channel. As shown in the table, we can obtain the positions of $X(3960)$ and $X_0(4140)$, on this basis, we can suggest that $X(3960)$ is a resonance state and $X_0(4140)$ is a virtual state~\cite{Suzuki:2009dc,Suzuki:2008rp}. This observation is consistent with the results shown in Fig.~\ref{fig2}, which clearly indicate that the formation of $X(3960)$ is primarily due to the interaction of $V_{D_s^+D_s^-,D_s^+D_s^-}$. Even without considering the contribution of $V_{D_s^+D_s^-,D_s^{\ast}\bar{D}_s^{\ast}}$ and $V_{D_s^{\ast}\bar{D}_s^{\ast},D_s^{\ast}\bar{D}_s^{\ast}}$, the state of $X(3960)$ can be understood as a bound state of $D_s^+D_s^-$. The behavior of the green dotted curves in Fig.~\ref{fig2}(a) further supports the notion that if $X_0(4140)$ is a virtual state, the contribution of $V_{D_s^{\ast}\bar{D}_s^{\ast},D_s^{\ast}\bar{D}_s^{\ast}}$ is weak.\\

In order to investigate the threshold effect of the kinematic effect in the vicinity and determine whether the $X(3960)$ peak structure is solely caused by the $D_s\bar D_s$ threshold, we disabled the coupled-channel effect, equivalent to directly finding the monocyclic graph contribution of the fig~\ref{fig1}. The data was then re-fitted, as shown in the table. Although the overall change in $\chi^2$ is small, it is evident that the height of the first peak undergoes a significant change, and the change in $\chi^2$ is primarily due to this peak. Therefore, it can be concluded that the pure kinematic effect alone is insufficient to form a peak structure.  The peak structure should indicate a state that actually exists.\\

	\section{Conclusion}
	\label{IV}

We analyze the observations of the LHCb Collaboration on the decay process $B^+\to D_s^+D_s^-K^+$. Our default model fits at the same time the $M_{D_s^+D_s^-}$, $M_{D_s^+K^+}$, and $M_{D_s^-K^+}$ of these three different invariant mass distributions, and the final fitting quality is $\chi^2/ndf\simeq1.42$. Without adding resonance states directly, we search the poles of $X(3960)$ and $X_0(4140)$ by coupled-channel model and finally determine the positions of $X(3960)$ and $X_0(4140)$. From this, we suggest that $X(3960)$ may be a resonance state and $X_0(4140)$ may be a virtual state. The determination of the pole positions is meaningful, which provides information for the research of $X(3960)$ and $X_0(4140)$, and does not lay a certain foundation for the study of their properties in the future. By turning off the coupled-channel effect and fitting the data again, we find that the overall fitting quality does not change very much, but from the final fitting result, it can be seen that the influence is relatively large at the position of $X(3960)$, and almost all the changes of $\chi^2$ come from the $X(3960)$ peak. Therefore, we suggest that the pure kinematic effect was not enough to form the $X(3960)$ peak structure. This conclusion will provide certain reference value for future research.\\ 

	\section{Acknowledgments}
	\label{V}

	This work is partly supported by the National Natural Science Foundation of China under Grants No. 12205002,
	partly supported by the the Natural Science Foundation of Anhui Province (2108085MA20,2208085MA10), 
	and partly supported by the key Research Foundation of Education Ministry of Anhui Province of China (KJ2021A0061).		

\bibliography{ref}	
\end{document}